# A Framework for Conceptualizing Islamic Bank Socialization in Indonesia

Suryo Budi Santoso[1] and Herni Justiana Astuti[2]

**Universitas Muhammadiyah Purwokerto, Purwokerto, Indonesia**
[1]suryobs@gmail.com, [2]herni99@gmail.com

**Abstract.** The purpose of this study is the design model of Islamic bank socialization in terms of four pillars (Business Institution, Formal Education, Islamic Scholar and Higher Education) through Synergy and Proactive. The location of the study was conducted in the Regency of Banyumas, Indonesia. The results of the survey on respondents obtained 145 respondents' answers that deserve to be analyzed. Data were analyzed using SEM models with Partial Least Squares approach, designing measurement models (outer models) and designing inner models. The results of the calculation outside the model of all measurements are more than the minimum criteria required by removing Formal Education from the model because it does not meet the requirements. While the inner model results show that the socialization model was only built by the Business Institution, Islamic Scholar and Higher Education through synergy and proactivity. All independent variables directly influence the dependent variable, while the intervening variables also significantly influence except the relationship of Islamic Scholar Islamic to Bank Socialization through Proactive.

**Keywords:** Islamic Bank, Socialization, Business Institution, Formal Education, Islamic Scholar, Higher Education

**Introduction**
Islamic banks in Indonesia established since 27 years ago. The development of assets of Islamic banks is slow. The percentage of Islamic bank assets compared to conventional banks is 3.93% [1]. The percentage of Islamic bank assets for 27 years is still small. Indonesia Islamic bank participation in the world is 5.5 percent in year 2014-2015 [2]. This figure is relatively small when compared to Bahrain and Malaysia that have reached 27.7 percent and 20.7 percent respectively [2].
This gives a slow signal for the development of Islamic banks in Indonesia. One important thing in developing Islamic banks in Indonesia is the design of socialization of Islamic bank. Islamic banks in Indonesia are a relatively new banking system when compared with conventional banking system. A good Islamic bank socialization design is expected to increase the interest of Indonesian people using Islamic banking services in everyday life. Muslims use the services of Islamic banks is part of worship in terms of economic activity in accordance with Islamic religion. Furthermore, Indonesia has the largest population of Muslims in the world (12,9 percent) which is a good opportunity for the development of Indonesian Islamic banks [3].
Kuwamura found that proactive is necessary in realizing the internationalization of education in Japan through various means such as increasing diversity effectively, developing different intercultural competencies. In other words Kawamura has found that proactive is needed to optimize results with various paths [4]. Socialization of Islamic banks is also required proactive. Therefore, as found by Goerdel states that proactive is very important for the successful achievement of management objectives. Managers who always make contact with network actors with the network often experience success than those who do not make contact. The proactive become important in the successful achievement of organizational or organizational goals [5]. A holistic perspective on how companies can be more proactive in their journey to becoming more sustainability orientated [6]. The circular business models imply significant challenges to proactive uncertainty reduction for the entrepreneur [7]. Proactive is important in business, companies, organization and also for uncertainty reduction.

The synergy between business institutions, formal education, Islamic scholars, and universities is necessary in the socialization of Islamic banking. Synergy is needed to generate organizational goals. Synergy can support dynamic change [8]. The socialization of Islamic finance is a joint responsibility between Bank Indonesia and the Financial Services Authority and Islamic financial practitioners. Strategic steps in the socialization of Islamic banks include targeting, integration, understanding, and implementation [9].
Therefore, as found by Goerdel states that proactive is very important for the successful achievement of management objectives. Managers who always make contact with network actors or synergize with the network often experience success than those who do not make contact. Thus, synergies and proactive become important in the successful achievement of organizational or organizational goals. Socialization is a process of cultural reconstruction by which individuals in each new generation are guided to construct some semblance of cultural continuity [10].
Literature Review
Research on the socialization of Islamic banks has not been widely carried out at the international level. There have been studies on sharia bank socialization in Indonesia, although the numbers are not large. Some researchers like the results of a national survey by the Authority of the literacy and inclusion of Indonesian Sharia banks were 6.63% and 9.61% in 2016 [11]. It means that people have started to use Islamic banking and financial products, but not many people understand about Islamic banking and financial products. So, the socialization about Islamic bank is important, including the products. Other researcher, Sakinah found that Islamic scholars have strategic role in the socialization of sharia banking because the ulama are as heirs of the prophets with the position and predicate as ahlul ilmi (expert) of science, ahlul khasyah (one who is taqwa), ahlul bashirah (has competence). The Islamic scholar or Ulama in Indonesia is one of strategies way to help the Islamic bank Socialization well.
The socialization of Islamic banks is currently implemented by some Islamic institutions in Indonesia. However, it is not appreciated to be so effective [12]. The socialization of Islamic banks is not effective in terms of: (1) socialization only in urban areas; (2) the public does not know and clearly understand Islamic financial institutions, but the insights and knowledge of Islamic banking are generally limited among academics and practitioners [13].

**Research Method**
Data obtained from questionnaires on people once associated transactions with the Islamic bank. The results of the survey on respondents obtained 145 respondents' answers are feasible to be analyzed. Then, the data were analyzed using SEM model with Partial Least Squares approach. In the analysis stage, there are two separate but sequential relationships. First, we must design the measurement model (outer model) to know the validity and reliability of indicators of latent variables. Second, the structural model is tested by designing an inner model.

Outer models are designed using multiple measurements. First, convergent validity or loading factor (minimum 0.5 is acceptable if research is still in the early stages of developing a scales measurement) [14]. Second, the average variance extracted is > 0.5 [15]. The next evaluation is composite reliability (a value > 0.6 indicates that the construction is reliable) [16].

## 4. Results And Discussion

### 4.1 Outer Model
The measurement model can proceed to the inner model stage if it meets valid and reliable criteria. It can be seen that table 2 illustrates the measurement model. All indicators on running 1 cannot be accepted because some loading factors are smaller than 0.5. These indicators are FE2 (0.436) and FE3 (-0.173). After the three indicators are removed from the model (see loading factors running 2), the results are obtained that all indicators can be accepted, except the FE1 indicator which has a perfect score. Thus, the Formal Education variable must be excluded from the model, while AVEs for the other variables already meets the requirements for values above 0.50. Then, all indicators and variables are valid. The model is said to be reliable if the value of Composite Reliability is more than 0.6. If seen in table 2, all variables have met reliable criteria.

Table 2. Variables, Indicators, Loading Factor, AVE and Composite Reliability

| Variables | Indicators | Loading Factors[a] Running 1 | Loading Factors[a] Running 2 | AVEs[b] | Composite Reliability[c] |
|---|---|---|---|---|---|
| Business Institution | BI1: Islamic Religious Education | 0.777 | 0.778 | 0.551 | 0.831 |
| | BI2: exemplary practice of educators | 0.772 | 0.772 | | |
| | BI3: Islamic education institutions | 0.740 | 0.740 | | |
| | B14: Islamic Business Institutions | 0.676 | 0.676 | | |
| Formal Education | FE1: Socialization at high schools | 0.905 | 1.000 | Rejected | Rejected |
| | FE2: Socialization at junior high schools | 0.436 | Rejected | | |
| | FE3: Socialization at elementary schools | -0.173 | Rejected | | |
| Islamic Scholar | IS1: the National Sharia Council | 0.795 | 0.795 | 0.637 | 0.875 |
| | IS2: speech of the Islamic scholars | 0.810 | 0.809 | | |
| | IS3: The material Islamic Scholar | 0.817 | 0.817 | | |
| Higher Education | HE1: Socialize Islamic banking products | 0.867 | 0.868 | 0.697 | 0.821 |
| | HE2: Socialization to Higher Education | 0.801 | 0.800 | | |
| Synergy | SY1: synergy with existing media | 0.753 | 0.755 | 0.582 | 0.848 |
| | SY2: synergy with community leaders | 0.796 | 0.796 | | |
| | SY3: synergy with Islamic scholar | 0.802 | 0.801 | | |
| | SY4: synergy with educator | 0.696 | 0.695 | | |
| Proactive | P1: The government proactively | 0.887 | 0.894 | 0.719 | 0.836 |
| | P2: Islamic banking proactively | 0.809 | 0.800 | | |
| Islamic Bank Socialization | IBS1: continuous learning process | 0.857 | 0.857 | 0.637 | 0.875 |
| | IBS2: transfer the understanding | 0.795 | 0.795 | | |
| | IBS3: dissemination of information | 0.802 | 0.803 | | |
| | IBS4: communication to the public | 0.733 | 0.733 | | |

AVE: Average Varian Extract
[a]Acceptable value of Loading Factor is greater than 0.5
[b]Acceptable value of AVE is greater than 0.5
[c]Acceptable value of Composite Reliability is greater than 0.6

### 4.2 Inner Model
Evaluation of structural models or inner models aims to predict relationships between latent variables. The inner model is evaluated by looking at the value of R-Square for latent endogenous constructs. Based on Figure 1, it can be seen that Synergy can explain the relationship with Business Institution, Islamic Scholar and Higher Education of 47.9%. Whereas Business Institution, Islamic Scholar and Higher Education can explain the relationship with Proactive by 42.9%. Islamic Bank Socialization variability explained by Synergy and Proactive by 49%.

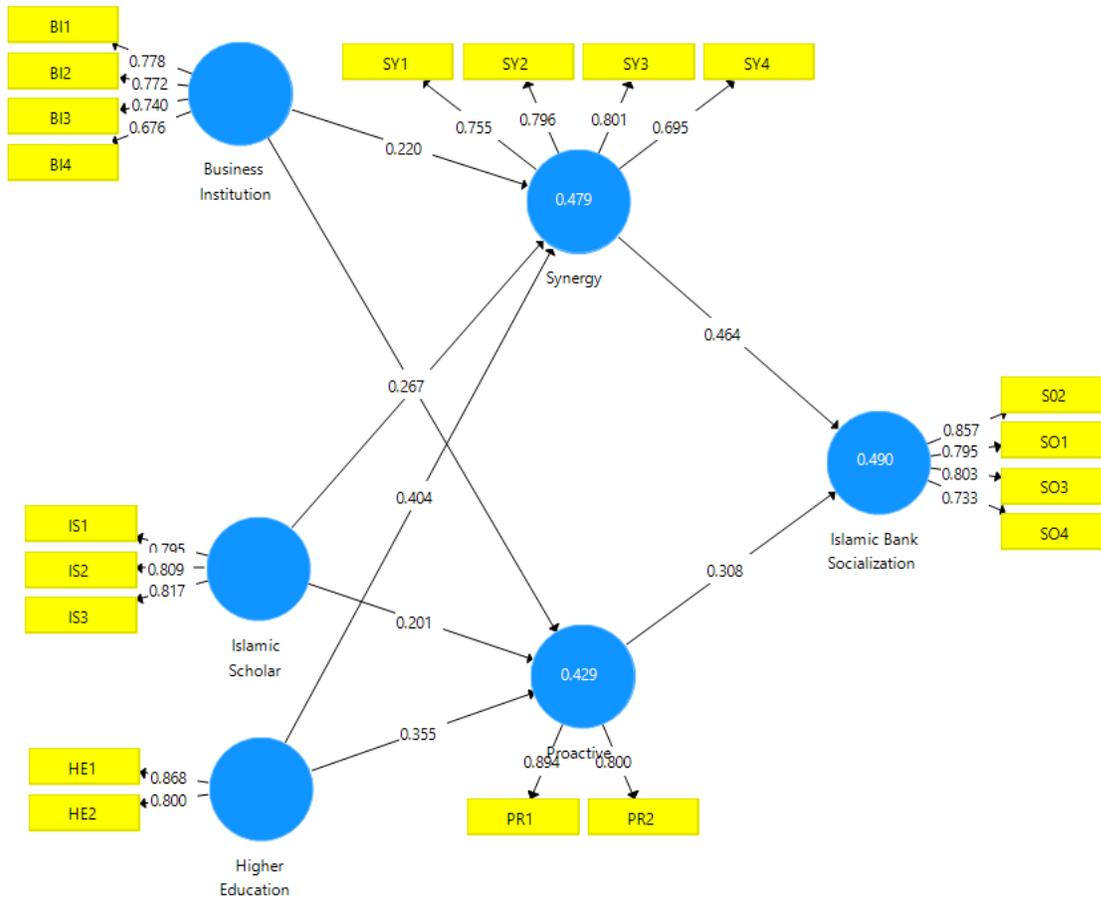

**Model Figure 1. Framework of Islamic Bank Socialization**

Table 3 explains the results of the direct relationship, path coefficient, and P value. There are 8 relationships of direct influence. All of the direct relationships have a significant effect, and also the direction of the relationship is positive.

**Table 3. Direct Relationships, Path Coefficients, P Values, Results**

| Direct Relationships | Path Coefficients | P values | Results |
|---|---|---|---|
| Business Institution ➔ Synergy | 0.220 | 0.008 | Accepted |
| Islamic Scholar ➔ Synergy | 0.247 | 0.013 | Accepted |
| Higher Education ➔ Synergy | 0.404 | 0.002 | Accepted |
| Business Institution ➔ Proactive | 0.267 | 0.004 | Accepted |
| Islamic Scholar ➔ Proactive | 0.201 | 0.031 | Accepted |
| Higher Education ➔ Proactive | 0.355 | 0.009 | Accepted |
| Synergy ➔ Islamic Bank Socialization | 0.464 | 0.000 | Accepted |
| Proactive ➔ Islamic Bank Socialization | 0.308 | 0.000 | Accepted |

Table 4 explains the results of the indirect relationship, path coefficient, and P value.
There are 6 relationships of indirect influence. All of the indirect relationships have a significant effect, and also the direction of the relationship is positive, except the relationship of Islamic Scholar Islamic to Bank Socialization through Proactive.

**Table 4. Indirect Relationships, Path Coefficients, P Values, Results**

| Direct Relationships | Path Coefficients | P values | Results |
|---|---|---|---|
| Business Institution ➔ Synergy ➔ Islamic Bank Socialization | 0.109 | 0.014 | Accepted |
| Business Institution ➔ Proactive ➔ Islamic Bank Socialization | 0.082 | 0.027 | Accepted |
| Islamic Scholar ➔ Synergy ➔ Islamic Bank Socialization | 0.115 | 0.022 | Accepted |
| Islamic Scholar ➔ Proactive ➔ Islamic Bank Socialization | 0.062 | 0.075 | Rejected |
| Higher Education ➔ Synergy ➔ Islamic Bank Socialization | 0.188 | 0.018 | Accepted |
| Higher Education ➔ Proactive ➔ Islamic Bank Socialization | 0.109 | 0.036 | Accepted |

## 4.3 Discussion

**a.) Effect Main Pillars to Socialization through Synergy**

Stakeholders synergize with business institutions, higher education and Islamic scholar to optimize Islamic bank socialization. Stakeholders consist of institutions of Islamic banks, government, and society. Business institutions, higher education and Islamic scholar to optimize Islamic banking socialization cannot be used directly. They should synergize to obtain maximum results in the socialization of Islamic banks. Business institutions that can be synergized are schools, madrassas, and pesantren (Islamic Boarding Schools). Madrasah or madrasas are a type of religious or specific school for Islamic studies, although this may not be the only topic being studied as well as the modern curriculum, while pesantren or Pondok Pesantren are Islamic boarding schools in Indonesia. School is clearly many learners are Muslim because as a country whose population of Islam must synergize. Madrasahs and Islamic Boarding Schools are almost certain that their students are Muslims. Thus, synergizing Madrasah and pesantren to socialize about Islamic banks really need to be done seriously. Furthermore, teachers and schools can set an example by using the services of an Islamic bank. This makes the students more confident because it has been given an example by teachers and the school. Moreover, Muslim businessmen use Islamic banking services in their daily activities. As a Muslim entrepreneur must have a commitment in socializing Islamic banks. Higher education is very important to be invited to work together in Islamic banking socialization. Through the three pillars of higher education, from teaching, research, and the community services certainly has a strategic aspect to be developed. Socialization can synergize more strongly through the strategic role of universities for the progress of Islamic banks in Indonesia. Finally, the connection with synergy is the Islamic Scholar. The role of Islamic scholar in the socialization of Islamic banks is very important. Islamic Scholar is very influential to his followers. It is this strategic thing that needs to be synergized well. If not done well, then followers of Islamic Scholars also tend not to use services from Islamic banking. As informed above that the Islamic Scholar is the variable that has the greatest influence on synergy with Islamic Scholar who disseminate information about the fatwa of scholars of Indonesia on Islamic banks as the biggest indicator of loading factor.

**b) Effect Main Pillars to Socialization through Proactive.**

Stakeholders must proactive with formal education, business institutions, higher education and Islamic scholar to optimize Islamic bank socialization. Islamic bank socialization requires proactive from stakeholders. Stakeholders must be proactive with; Formal education, business institutions, higher education and Islamic scholar. Proactive needs more items than synergy. With the proactive of the four items above, then the socialization of Islamic banks can be better. Proactive with formal education, especially junior high and high school, then Islamic bank socialization will be more effective. The results show that both above can be accepted as an item that needs to be done proactively.

c) Socialization of Islamic banks requires synergy and proactive from business institutions, higher education and Islamic scholar to succeed. Socialization of Islamic banks is not synergistic and not proactive with formal education, especially at the elementary level and junior high school. Formal education can still be synergistic and proactive in senior high schools in relation o the socialization of Islamic banks. This can happen because the respondent assumes that the students have not understood the Islamic Bank's material in depth, especially for students of grade one to four, may be understood for the fifth and sixth grade.

## 5. Conclusions

To design an optimum model of socialization of Islamic banks to the public at large by synergizing the four main pillars of socialization namely business institutions, higher education and Islamic scholars. Formal education involving elementary, secondary and high schools has no effect on the optimization of socialization. Formal education has only an indirect effect on the optimization of socialization through proactive. However, not all levels in formal education become the raw material for the socialization process. Basic education does not include indicators that build socialization model. By the respondents considered elementary school students still do not understand the existence of Islamic banks. This limitation can be a suggestion for further research to pay more attention to the study of socialization of Islamic banks that are appropriate for elementary school students. In addition, it is necessary to design an integrative Islamic Scholar speech on the fatwa of Indonesia Islamic Scholar, as it is the largest indicator affecting the success of socialization through synergy.